\documentstyle[aps,prl,twocolumn,epsf]{revtex}
\newcommand{\be}{\begin{equation}}
\newcommand{\ee}{\end{equation}}
\newcommand{\bea}{\begin{eqnarray}}
\newcommand{\eea}{\end{eqnarray}}
\newcommand{\nn}{\nonumber}

\def\bfnabla{\mbox{\boldmath $\nabla$}}

\def\lQ{\Lambda_{\rm QCD}}

\def\als{\alpha_{\rm s}}
\def\siml{{\ \lower-1.2pt\vbox{\hbox{\rlap{$<$}\lower6pt\vbox{\hbox{$\sim$}}}}\ }}

\begin{document}
\twocolumn[\hsize\textwidth\columnwidth\hsize\csname
@twocolumnfalse\endcsname\preprint{}
\title{\bf New predictions for inclusive heavy-quarkonium P-wave decays}
\author {Nora Brambilla$^1$, Dolors Eiras$^2$, Antonio Pineda$^3$, Joan
  Soto$^2$ and Antonio Vairo$^4$}
\address{$^1$ INFN and Dipartimento di Fisica dell'Universit\`a di Milano \\
  via Celoria 16, 20133 Milan, Italy}
\address{$^2$ Dept. d'Estructura i Constituents de la Mat\`eria and IFAE,
  U. Barcelona \\ Diagonal 647, E-08028 Barcelona, Catalonia, Spain}
\address{$^3$ Institut f\"ur Theoretische Teilchenphysik, U. Karlsruhe,
  D-76128 Karlsruhe, Germany}
\address{$^4$ Theory Division CERN, 1211 Geneva 23, Switzerland}
\maketitle\begin{abstract}
\noindent
We show that some NRQCD colour-octet matrix elements can be written in terms of
(derivatives of) wave functions at the origin and non-perturbative universal
constants once the factorization between the soft and ultrasoft scale is
achieved by using an effective field theory where only ultrasoft degrees
of freedom are kept as dynamical entities. This allows us to derive a new set of relations
between inclusive heavy-quarkonium P-wave decays into light hadrons with different principal
quantum number and with different heavy flavour. In particular, we can estimate
the branching ratios of bottomonium P-wave states by using charmonium data.
\end{abstract}\vskip 2pc]

Inclusive P-wave decays to light hadrons have proved to be an optimal testing
ground of our understanding of heavy quarkonia.
The use of NRQCD \cite{BBL0,BBL} allowed a description of these decays in terms
of expectation values of some 4-heavy-quark operators at a quantum-field
level in a systematic way. Besides the so-called colour-singlet operators, for which their
expectation values could be related to wave functions in an intuitive way,
there were also colour-octet operators. The latter were decisive in solving
the infrared sensitivity of earlier calculations \cite{earlier}.
It has been thought so far that these colour-octet expectation values could not
be related with a Schr\"odinger-like formulation in any way.

We show in this letter that it is not so. For certain states, the expectation values of
colour-octet operators can also be written in terms of
wave functions and additional bound-state-independent non-perturbative
parameters. We shall focus on the operators relevant to P-wave decays
into light hadrons, but it should become apparent that this is a general feature.

The line of developments that has led us to this result is the following.
It was pointed out in Ref. \cite{Mont} that NRQCD still contains
dynamical scales, which are not relevant to the kinematical situation
of the lower-lying states in heavy quarkonium
(energy scales larger than the ultrasoft scale: $m v^2$,
$v$ being the relative velocity of the heavy quark and $m$ its mass).
Hence, further simplifications occur if we integrate them out.
We call pNRQCD the resulting effective field theory (as in \cite{M1}; note that
in \cite{long}, in the situation $\lQ \gg mv^2$, the EFT was called pNRQCD$^\prime$).
When the typical scale of non-perturbative physics, say $\lQ$, is
smaller than the soft scale $mv$, and larger than the ultrasoft scale $m v^2$,
the soft scale can be integrated out perturbatively. This leads to
an intermediate EFT that contains, besides the singlet,
also octet fields and ultrasoft gluons as dynamical degrees of freedom \cite{Mont,long}.
These are eventually integrated out by the (non-perturbative)
matching to pNRQCD \cite{long}. When $\lQ$ is of the order of the soft scale,
the (non-perturbative) matching to pNRQCD has to be done in one single step.
This framework has been developed in a systematic way in Ref. \cite{M1}.

In this letter we will compute the inclusive P-wave decay widths into
light hadrons at leading order for $\lQ \gg mv^2$ by using pNRQCD.  In
this situation the singlet is the only dynamical field in pNRQCD (Goldstone
bosons are also present, but they play a negligible role in the present analysis
and will be ignored), if hybrids and other degrees of freedom associated
with heavy--light meson pair threshold production develop a mass gap
of ${\mathcal O}(\lQ)$, as we will assume in what follows \cite{long,M1},
or if they play a minor role in the heavy-quarkonium dynamics.
Therefore, the pNRQCD Lagrangian reads \cite{long,M1}
\begin{eqnarray}
& & {\mathcal{L}}_{\rm pNRQCD}=
{\rm Tr} \,\bigg\{ {\rm S}^\dagger \left( i\partial_0 - h  \right) {\rm S} \bigg \} \, ,
\label{lpnrqcd}
\end{eqnarray}
where $h$ is the pNRQCD Hamiltonian, to be determined by matching the EFT to NRQCD.
The total decay width of the singlet heavy-quarkonium state is then given by
\be
\Gamma= - 2\, {\rm Im} \, \langle n,L,S,J| h  |n,L,S,J \rangle,
\ee
where $|n,L,S,J \rangle$ are the eigenstates of the Hamiltonian $h$.
The imaginary parts are inherited from the 4-heavy-fermion NRQCD Wilson
coefficients and, for P-wave decays, first appear as local (delta-like)
${\mathcal{O}}(1/m^4)$ potentials in the pNRQCD Lagrangian. The relevant
structure reads (we shall concentrate on potentials, which inherit imaginary parts
from the NRQCD operators and which contribute to P-wave states at first order in
quantum-mechanical perturbation theory (QMPT)):
\be
-2 \, {\rm Im} \, h \bigg|_{\rm P-wave}
=  \, F_{SJ}{\mathcal{T}}_{SJ}^{ij}\frac{{\bfnabla}^i_{\bf r}
\delta^{(3)}({\bf r}) {\bfnabla}^j_{\bf r}}{m^4} \,,
\label{hs}
\ee
where ${\mathcal{T}}_{SJ}^{ij}$ corresponds to the spin and total angular
momentum wave-function projectors. What is now left is to compute $F_{SJ}$,
i.e. to perform the matching between NRQCD and pNRQCD. For the situation
A): when $mv \gg \lQ \gg mv^2$, by taking the results of \cite{long}, and
for the more general situation B): when $\lQ \siml mv$, by using the formalism
of Refs. \cite{M1}. In both situations we get:
\be
F_{SJ} =  \! -2N_c \, {\rm Im} \, f_1(^{2S+1}{\rm P}_J)
- \! {4 T_F\over 9 N_c} \, {\mathcal E} \, {\rm Im} \, f_8(^{2S+1}{\rm S}_S),
\label{fsj}
\ee
where $f_1(^{2S+1}{\rm{L}}_J)$ and
$f_8(^{2S+1}{\rm{L}}_J)$ are the short-distance Wilson coefficients of NRQCD
as defined in Ref. \cite{BBL} and
\be
{\mathcal{E}} = \!  T_F \!\! \int_{0}^{\infty} \!\!\!\! d\tau \,\tau^3  \left\langle g {\bf E}^a(\tau,{\bf 0})
\Phi_{ab}(\tau,0;{\bf 0}) g{\bf E}^b(\tau,{\bf 0}) \right\rangle.\!\!
\label{ee}
\ee
{\bf \noindent A) \, P-wave potentials for $mv \gg \lQ \gg mv^2$.}\\
In this case the matching from NRQCD to pNRQCD at the scale $\lQ$ can be
done in two steps. In the first step, which can be done perturbatively, we
integrate out the scale $mv$ and end up with an EFT, which contains singlet ($\rm S$)
and octet ($\rm O$) fields as dynamical degrees of freedom.
At the next-to-leading order in the multipole expansion the Lagrangian reads \cite{Mont,long}:
\begin{eqnarray}
& &
{\cal L} =
{\rm Tr} \,\Big\{ {\rm S}^\dagger \left( i\partial_0 - h_s  \right) {\rm S} +
+ {\rm O}^\dagger \left( iD_0 - h_o   \right) {\rm O} \Big\} \nonumber \\
& & + {\rm Tr} \left\{\!
{\rm O}^\dagger {\bf r} \cdot g{\bf E}\,{\rm S}
+ \hbox{h.c.} +
{{\rm O}^\dagger {\bf r} \cdot g{\bf E} \, {\rm O} \over 2} +
{{\rm O}^\dagger {\rm O} {\bf r} \cdot g{\bf E} \over 2} \!\right\}
\nn \\
& & - {1\over 4} F_{\mu \nu}^{a} F^{\mu \nu \, a}\, ,
\label{pnrqcd0}
\end{eqnarray}
where $h_s$ and $h_o$ have to be determined by matching to NRQCD. They read as follows
\begin{eqnarray}
& & h_s = -{{\bfnabla}^2_{\bf r} \over m}+V_s(r)
+ \, ...  \nonumber \\
& & + \, N_c \, f_1(^{2S+1}{\rm{P}}_J)
{\mathcal{T}}_{SJ}^{ij}\frac{{\bfnabla}^i_{\bf r}
\delta^{(3)}({\bf r}) {\bfnabla}^j_{\bf r}}{m^4} + \, ..., \nonumber \\
& & h_o = -{{\bfnabla}^2_{\bf r}\over m}+V_o(r)
+ \, ...  \nonumber \\
& &
+ \, T_F \, f_8 (^{2S+1}{\rm{S}}_S){\mathcal{T}}_{S} \frac{\delta^{(3)}({\bf r})}{m^2} + \, ... \, ,
\end{eqnarray}
neglecting centre-of-mass recoil terms; ${\mathcal{T}}_S$ corresponds
to the total spin projector. Beyond
${\mathcal{O}}(1/m)^0$ we have only displayed the terms that are relevant to
our calculation. In the second step
we integrate out (non-perturbatively) the gluons and the octet field ending up
with the pNRQCD Lagrangian (\ref{lpnrqcd}). The Hamiltonian $h$ has to be
determined by matching the two effective field theories. It reads $h  = h_s
+\delta h_s$, with (at leading non-vanishing order in the multipole expansion)
\begin{eqnarray}
& & \delta h_s  = -i \frac{T_F}{N_c}
\int_0^{\infty} d\tau \, e^{ih_s\frac{\tau}{2}}  \left
< {\bf r}\cdot g{\bf E}^a (\tau,{\bf 0}) \right.
\nonumber \\
& &
\left. e^{-ih_o\tau} \, \Phi_{ab} (\tau, 0 ; {\bf 0})
\, g{\bf E}^b (0,{\bf 0}) \cdot {\bf r} \right > e^{ih_s \frac{\tau}{2}}\,  ,
\end{eqnarray}
where consistency with $\lQ \gg mv^2$ requires an expansion of the exponentials of
$h_o$ and $h_s$.  Taking into account that we are interested in P-wave states,
only the perturbation that puts one $h_o$ to each side of the
${\mathcal{O}}(1/m^2)$ S-wave potential survives at leading order.
The final result reads:
\begin{eqnarray}
& & \! {\rm Im} \, \delta h_s\bigg|_{\rm P-wave} \!\!\!\!\! = \frac{2T_F}{9 N_c}
{\mathcal{E}}
\, \frac{{{\bfnabla}_{\bf r}} \delta^{(3)}({\bf r})
{{\bfnabla}_{\bf r}}}{m^4} {\mathcal{T}}_S  \, {\rm Im} f_8(^{2S+1}{\rm{S}}_S)
\nn\\
\label{m2pwave1}
\end{eqnarray}
which plugged into Eq. (\ref{hs}) gives Eq. (\ref{fsj}).
This shows how a colour-octet operator in NRQCD becomes a colour-octet
potential in the EFT of Eq. (\ref{pnrqcd0}) and, eventually, contributes
to a colour-singlet potential in pNRQCD, which is one of our main points.

{\bf \noindent B) \, P-wave potentials for $\lQ \siml mv$.}\\
In the case $\lQ \siml mv$ the matching from NRQCD to pNRQCD at the scale
$\lQ$ has to be done directly, since no other relevant scales are supposed to lie
between $m$ and $mv$. The only dynamical degree of freedom of pNRQCD is the
heavy-quarkonium singlet field $\rm S$. The Lagrangian has been written in (\ref{lpnrqcd}).
The Hamiltonian $h$ is obtained by matching (non-perturbatively) to NRQCD,
order by order in $1/m$, within a Hamiltonian formalism \cite{M1}.
In this letter we only sketch the main steps of the derivation.
In short, we can formally expand the NRQCD Hamiltonian in $1/m$:
\be
H_{\rm NRQCD}= H_{\rm NRQCD}^{(0)} + {1\over m} H_{\rm NRQCD}^{(1)}+\cdots\,.
\ee
The eigenstates of the heavy-quark-antiquark sector can be labelled as
$$
|{\underline g};{\bf x}_1, {\bf x}_2\rangle=|{\underline g};{\bf x}_1,
    {\bf x}_2\rangle^{(0)}+{1 \over m}|{\underline g};{\bf x}_1, {\bf
        x}_2\rangle^{(1)}+\cdots ,
$$
where $\underline g$ labels the colour-related degrees of freedom
(we do not explicitly display spin labels for simpli\-ci\-ty). Assuming a mass gap
of ${\cal O}(\lQ)$ much larger than $mv^2$, all the excitations
(${\underline g}\not= {\underline 0}$) decouple and the ground state
(${\underline g}= {\underline 0}$) corresponds to the singlet state.
Therefore, the matching condition reads
\bea
&&
\langle {\underline 0};{\bf x}_1, {\bf x}_2|H|{\underline 0};{\bf
  x}'_1, {\bf x}'_2\rangle
\\
\nn
&&
=h({\bf x}_1, {\bf x}_2,\bfnabla_{{\bf x}_1},\bfnabla_{{\bf x}_2})
\delta^{(3)}({\bf x}_1-{\bf x}'_1)\delta^{(3)}({\bf x}_2-{\bf x}'_2).
\eea
Up to ${\cal O}(1/m^4)$ the imaginary contributions are only carried by the Wilson
coefficients of the dimension 6 and 8 4-heavy-fermion operators in NRQCD.
Since we are only interested in Eq. (\ref{hs}) a huge simplification occurs and only two
contributions survive. From the dimension  8 operators we obtain
\bea
\label{m4pwave2}&& {\rm Im} \, \delta h \,
\delta^{(3)}({\bf x}_1-{\bf x}'_1)\delta^{(3)}({\bf x}_2-{\bf x}'_2)
\\
\nn
&&
= {1 \over m^4} \,  {\rm Im} \,
^{(0)}\langle {\underline 0};{\bf x}_1, {\bf x}_2|H^{(4)}_{\rm NRQCD}|{\underline 0};{\bf x}'_1,
    {\bf x}'_2\rangle^{(0)} \bigg|_{\rm P-wave}
\\
\nn
&&=N_c  \, {\mathcal{T}}_{SJ}^{ij} \, {\rm Im} \, f_1(^{2S+1}{\rm P}_J) \,
\frac{{\bfnabla}_{{\bf r}}^i \delta^{(3)}({\bf r}) {\bfnabla}_{{\bf r}}^j}{m^4}
\\
\nn
&&
\qquad
\times \,\delta^{(3)}({\bf x}_1-{\bf x}'_1)\delta^{(3)}({\bf x}_2-{\bf x}'_2).
\eea
On the other hand, we also have contributions from the iteration of lower-order
$1/m$ corrections to the NRQCD Hamiltonian with the dimension 6 4-heavy-fermion
operators. The only term that contributes to Eq. (\ref{hs}) is
\bea
\label{m2pwave2}
&& {\rm Im} \, \delta h \, \delta^{(3)}({\bf x}_1-{\bf x}'_1)\delta^{(3)}({\bf x}_2-{\bf x}'_2)
\\
\nn
&&
= {1 \over m^4} \, {\rm Im} \,
{}^{(1)}\langle {\underline 0};{\bf x}_1, {\bf x}_2|H^{(2)}_{\rm NRQCD}|{\underline 0};{\bf x}'_1,
    {\bf x}'_2\rangle^{(1)} \bigg|_{\rm P-wave}.
\eea
The explicit computation of the right-hand side of Eq. (\ref{m2pwave2}) gives
(as far as the P-wave contribution is concerned)
Eq. (\ref{m2pwave1}). Therefore, the sum of the contributions from
Eqs. (\ref{m4pwave2}) and (\ref{m2pwave2}) coincides with Eq. (\ref{hs}),
after the replacements (\ref{fsj}) and (\ref{ee}). \vspace{0mm}\\

We can now obtain the decay widths by using Eq. (\ref{hs}). At first order in QMPT, we obtain:
\begin{eqnarray}
& & \Gamma (\chi^S_{QJ} (n{\rm{P}}) \rightarrow {\rm{LH}}) \! = \!
\Bigl [ \, \frac{3N_c}{\pi} \, {\rm Im}\,   f_1(^{2S+1}{\rm{P}}_J) + \frac{2
  T_F}{3\pi N_c} \Bigr. \nonumber \\
& & \Bigl. {\rm Im} \,  f_8(^{2S+1}{\rm{S}}_S) \, {\mathcal{E}}  \,
\Bigr ] \, \frac{| R'_{Q n 1} (0) |^2}{m^4} ,
\label{(10)}
\end{eqnarray}
where
$\chi^{1}_{Q J}(n {\rm{P}}):=\chi_{Q J}(n {\rm{P}})$,
$\chi^{0}_{Q J}(n {\rm{P}}):=h_{Q}(n {\rm{P}})$ ($Q=b,c$), $n$ is the principal quantum number,
and $R_{Q n 1} (r)$ the radial wave function at leading order.
Comparing with the standard NRQCD formula, where spin symmetry has already been used, we have:
\be
\langle h_Q(n{\rm{P}})\vert O_8(^1 {\rm{S}}_0)\vert h_Q(n{\rm{P}}) \rangle
(\mu)= \frac{\vert R^{\prime}_{Qn1}( 0)\vert^2} {3\pi N_c m^2}T_F {\mathcal{E}}(\mu).
\label{matrix}
\ee
The information gained with this formula is that all non-perturbative flavour
and principal quantum number dependence is encoded in the wave function, as
in the colour-singlet operators. The additional non-perturbative parameter
${\mathcal{E}}(\mu)$ is universal: it only depends on the light degrees of
freedom of QCD. This implies that the following relation between decay widths
is also universal:
\begin{eqnarray}
\label{prediction}
& & {\mathcal{E}}(\mu)=- {9N_c^2\over 2T_F} \times   \\
\nn
& & { {\rm Im} \, f_1(^{2S+1}{\rm P}_J) \! - \! {\rm Im} \,
  f_1(^{2S^{\prime}+1}{\rm P }_{J^{\prime}})
{\Gamma (\chi^S_{Q J}(n{\rm{P}})\rightarrow {\rm{LH}})
\over \Gamma (\chi^{S^{\prime}}_{Q J^{\prime}}(n{\rm{P}})\rightarrow {\rm{LH}})}
\over
{\rm Im} \, f_8(^{2S+1}{\rm{S}}_S) \! - \! {\rm Im} \, f_8(^{2S^{\prime}+1}{\rm{S}}_{S^{\prime}})
{\Gamma (\chi^S_{Q J}(n{\rm{P}})\rightarrow {\rm{LH}})\over
\Gamma (\chi^{S^{\prime}}_{Q J^{\prime}}(n {\rm{P}})\rightarrow
{\rm{LH}})}}.
\end{eqnarray}

It is interesting to notice that the UV behaviour of ${\mathcal{E}}$ has the
logarithmic divergence:
\be
{\mathcal{E}}(\mu) \simeq 12\,N_c C_F\, {\als \over \pi}\,\ln \mu,
\ee
which matches exactly the IR log of the ${\mathcal{O}}(\alpha_s)$
correction of ${\rm Im}\,f_1(^{2S+1}{\rm{P}}_J)$, and hence the
cancellation originally observed in \cite{BBL0} is fulfilled. Then one could
consider the LL RG improvement of ${\mathcal{E}}$ by using the results of
Ref. \cite{BBL} for the running of the octet-matrix element. One obtains
($\beta_0 = 11 C_A/3 - 4n_fT_F/3$):
\be
{\mathcal{E}} (\mu)= {\mathcal{E}} (\mu')+
{24 N_c C_F\over \beta_0}\ln {\als(\mu')\over \als(\mu)}.
\label{running}
\ee

Let us apply the above results to actual quarkonium, under the assumption that
our framework, discussed in the paragraph before Eq. (\ref{lpnrqcd}),
provides a reasonable description for the P-wave states observed in nature.
The numerical extraction of ${\mathcal{E}}$ is a delicate task, since
several of the Wilson coefficients (see Ref. \cite{Petrelli} for a full
list of them) have large next-to-leading order contributions, which may spoil
the convergence of the perturbative series. This problem is not specific of
our formalism, but belongs to the standard formulation of NRQCD.
Here, in order to give an estimate, we only use those data that provide
more stable results in going from the leading to the next-to-leading order,
more precisely the average of Eq. (\ref{prediction}) for $(J,S)=(1,1)$, $(J',S')=(0,1)$,
and $(J,S)=(1,1)$, $(J',S')=(2,1)$. The experimental data have been
taken from \cite{pdg} and updated accordingly to \cite{amb00,mussa}.
Our final value reads
\begin{eqnarray}
{\mathcal{E}} (1\,{\rm GeV}) = 5.3^{+3.5}_{-2.2}\hbox{(exp)},
\label{numEE}
\end{eqnarray}
where we have used the NLO results for the Wilson coefficients with a LL improvement.
The errors only refer to the experimental uncertainties on the decay widths.
Theoretical uncertainties mainly come from subleading operators in the power counting
(${\cal O}(v)$ suppressed) and subleading terms in the perturbative expansion
of the Wilson coefficients (${\cal O}(\als)$ suppressed), whose
bad convergence may affect considerably the figure of Eq. (\ref{numEE}).
We feel, therefore, that further studies, maybe along the lines of Refs. \cite{chen},
are needed before a complete numerical analysis, including theoretical
uncertainties, can be done. In any case, the above figure is compatible with
the values that are usually assigned to the NRQCD octet and
singlet matrix elements (e.g. from the fit of \cite{Maltoni}
one gets ${\mathcal{E}} (1\, {\rm GeV}) = 3.6^{+3.6}_{-2.9} \hbox{(exp)}$).
The above figure is also compatible with the charmonium (quenched) lattice data of
\cite{latbo}, whereas, if the running (\ref{running}) is taken into account,
bottomonium lattice data, quenched \cite{latbo} and unquenched \cite{latbo2},
appear to give a lower value.
Note that, in the language of Refs. \cite{latbo,latbo2}, Eq. (\ref{matrix}) reads
${\cal E}(\mu) = 81 \, m_{b,c}^2 {\cal H}_8(\mu)/{\cal H}_1|_{b,c}$, which implies
${\cal H}_8(\mu)/{\cal H}_1|_{b} \times {\cal H}_1/{\cal H}_8(\mu)|_{c} =  m_c^2/m_b^2$.
For all quarkonium states that satisfy our assumptions this
equality must be fulfilled by lattice results for any number of
light fermions and for any value of the heavy-quark masses.

By using the estimate (\ref{numEE}),
we can also predict the branching ratios for the $n=1,2$ P-wave bottomonium states. We obtain
\be
\!\!
{\Gamma (\chi^1_{b1} (1{\rm{P}})) \over \Gamma (\chi^1_{b2} (1{\rm{P}}))} =
{\Gamma (\chi^1_{b1} (2{\rm{P}})) \over \Gamma (\chi^1_{b2} (2{\rm{P}}))} =
0.50^{+0.06}_{-0.04},
\ee
where only the errors inherited from Eq. (\ref{numEE}) have been included.
For what concerns theoretical uncertainties, the comments after Eq. (\ref{numEE}) apply also
here (with a better behaviour of the perturbative series).
Note that the first equality holds independently from Eq. (\ref{numEE}) and from the use
of charmonium data and, hence, provides a more robust prediction.
The remaining branching ratios can be obtained using spin symmetry.
Notice also that, although no model-independent predictions can be
made for the decay widths (they depend on the wave function at the origin,
which is flavour and state dependent), our results allow any model that
gives a definite value to $R^{\prime}_{Qn1}(0)$ to make definite predictions.

In conclusion, we have exploited the fact that NRQCD still contains irrelevant degrees of
freedom for certain heavy quarkonium states, which can be integrated out in
order to constrain the form of the matrix elements of colour-octet operators.
We have focused on the operators relevant to P-wave decays, which allowed us
to produce concrete, new, rigorous results. However, it should be clear
from the structure of the pNRQCD Lagrangian itself, that similar results can
be obtained for matrix elements of any colour-octet operator.  \vspace{0mm}\\

N.B. thanks R. Mussa for an illuminating correspondence. N.B. and A.V. thank
G. Bali and \- C. Davies for discussions. D.E. and J.S. are supported by the
AEN98-031 (Spain) and the 1998SGR 00026 (Catalonia). D.E. also acknowledges
financial support from a MEC FPI fellowship (Spain). A.P. thanks the
University of Barcelona and CERN for hospitality while part of this work was
carried out. A.V. thanks F. Maltoni for providing him with the second of
Refs. \cite{Petrelli}. A.V. is supported by the European Community through the
Marie-Curie fellowship HPMF-CT-2000-00733.
\vspace{-0mm}\\

\end{document}